\documentclass[aps,pre,groupedaddress]{revtex4}

\usepackage{graphicx}
\usepackage{float}
\usepackage{mathrsfs}
\usepackage{amsmath,amssymb}
\usepackage{hhline}
\usepackage{dcolumn}
\usepackage{bm}

\newcommand{\be}{\begin{equation}}
\newcommand{\ee}{\end{equation}}
\newcommand{\ba}{\begin{eqnarray}}
\newcommand{\ea}{\end{eqnarray}}

\begin{document}
\title{Critical and Nonpercolating Phases in Bond Percolation on the Song--Havlin--Makse Network}
\author{Kazuki Wataya}
\affiliation{Graduate School of Science and Engineering, Ibaraki University, 2-1-1, Bunkyo, Mito, 310-8512, Japan}
\author{Takehisa Hasegawa}
\email{takehisa.hasegawa.sci@vc.ibaraki.ac.jp}
\affiliation{Graduate School of Science and Engineering, Ibaraki University, 2-1-1, Bunkyo, Mito, 310-8512, Japan}

\begin{abstract}
We investigate bond percolation on the Song--Havlin--Makse (SHM) network, a scale-free tree with a tunable degree exponent and dimensionality.
Using a generating function approach, we analytically derive the average size and the fractal exponent of the root cluster for deterministic cases.
Our analysis reveals that bond percolation on the SHM network remains in a nonpercolating phase for all $p < 1$ when the network is fractal (i.e., finite-dimensional), whereas it exhibits a critical phase, where the cluster size distribution follows a power-law with a $p$-dependent exponent, throughout the entire range of $p$ when the network is small-world (i.e., infinite-dimensional), regardless of the specific dimensionality or degree exponent.
The analytical results are in excellent agreement with Monte Carlo simulations.
\end{abstract}

\maketitle

\section{Introduction}

Percolation processes on {\it complex networks}~\cite{newman2018networks} have been widely used to model failures, attacks, and information propagation~\cite{li2021percolation}. 
In bond percolation, each edge of a network is retained with probability $p$ and removed with probability $1 - p$. 
As $p$ increases from 0, the system typically exhibits a phase transition at a critical point $p_c$, separating a nonpercolating phase and a percolating phase~\cite{stauffer2018introduction}. 
These phases are characterized by the percolation order parameter $s_{\mathrm{max}}(p)$, defined as the fraction of nodes in the largest cluster formed by the retained edges.
In the large size limit, i.e., when the number $N$ of nodes in the network approaches infinity, $s_{\mathrm{max}}(p) = 0$ in the nonpercolating phase, indicating that all clusters are finite, while $s_{\mathrm{max}}(p) > 0$ in the percolating phase, signifying the emergence of an infinite cluster, often called the giant component. 
The critical point $p_c$ and associated critical exponents depend sensitively on the structure of the underlying network~\cite{dorogovtsev2008critical}. 
Over the past two or three decades, considerable research has focused on how structural heterogeneities of complex networks affect percolation critical properties~\cite{li2021percolation}. 
A prominent finding concerns the {\it scale-free} (SF) property~\cite{barabasi1999emergence}, where the degree distribution follows $P(k) \propto k^{-\gamma}$.
The degree exponent $\gamma$ characterizes the degree heterogeneity of SF networks: smaller values of $\gamma$ result in heavier-tailed degree distributions.
For $\gamma \leq 3$, the critical point vanishes ($p_c=0$), implying exceptional robustness of SF networks against random failures~\cite{albert2000error,cohen2000resilience,callaway2000network}. 
Beyond degree heterogeneity, the impacts of clustering and degree correlations on percolation transitions have also been extensively studied~\cite{li2021percolation}.

It is known that an atypical percolation transition, which is absent in homogeneous systems, can occur on nonamenable graphs~\cite{benjamini1996percolation}.  
A nonamenable graph is defined as a transitive infinite graph with a positive Cheeger constant. 
Typical examples include regular trees and hyperbolic lattices. 
A striking feature of percolation on such graphs is the existence of an intermediate phase, hereafter referred to as the {\it critical phase}, which lies between the nonpercolating and percolating phases.
Mathematically, this phase is characterized by the emergence of infinitely many infinite clusters~\cite{benjamini1996percolation,benjamini2001percolation}.  
From the perspective of statistical physics, Monte Carlo simulations performed  on the enhanced binary tree were used to characterize this atypical percolation transition~\cite{nogawa2009monte,baek2009comment,nogawa2009reply}.  
Bond percolation on the enhanced binary tree exhibits two critical points, $p_{c1} \approx 0.306$ and $p_{c2} \approx 0.564$~\cite{nogawa2009monte}, defining three distinct phases: nonpercolating phase ($0 \le p < p_{c1}$), critical phase ($p_{c1} < p < p_{c2}$), and percolating phase ($p_{c2} < p \le 1$). 
In the critical phase, the cluster size distribution $n(s)$---defined as the average number of clusters of size $s$ per node---follows a power law, $n(s) \propto s^{-\tau(p)}$, with exponent $\tau(p)$ varying with $p$. 
As will be shown later, these phases are more appropriately characterized by the fractal exponent $\psi(p)$ rather than the conventional order parameter $s_{\mathrm{max}}(p)$~\cite{nogawa2009monte,hasegawa2013profile,hasegawa2014critical}.  
Following this line of research, the critical phase has also been observed in various types of inhomogeneous networks~\cite{boettcher2009patchy,hasegawa2010generating,hasegawa2010critical,boettcher2012ordinary,hasegawa2012phase,hasegawa2013hierarchical,nogawa2014transition,singh2014scaling,boettcher2015classification,nogawa2016local}.
Nevertheless, the fundamental question remains unanswered: what structural properties of inhomogeneous networks give rise to the critical phase?

Trees---networks without loops---are among the simplest structures known to exhibit a critical phase.  
For example, binary trees lack the percolating phase and exhibit only the nonpercolating and critical phases, with critical points at $p_{c1} = 1/2$ and $p_{c2} = 1$~\cite{hasegawa2014critical}.  
While percolation on regular trees and Galton--Watson trees have been extensively studied, investigations into the critical phase on inhomogeneous trees remain limited.  
An exception is a study~\cite{hasegawa2010critical} reporting that bond percolation on growing trees with a preferential attachment mechanism exhibits a critical phase throughout the entire range of $p$, regardless of the degree exponent $\gamma$.
Further research is needed to clarify how the inhomogeneous structure of trees influences the nature of percolation transition.

The Song--Havlin--Makse (SHM) network~\cite{song2006origins} is recursively constructed, admits renormalization techniques, and realizes a tree with a power-law degree distribution, while having the advantage that its dimensionality can be continuously tuned from finite (fractal) to infinite (small-world). 
Previous studies of bond percolation on hierarchical small-world networks, where renormalization techniques are applicable, have reported that the system exhibits a critical phase and a percolating phase, but not a nonpercolating phase~\cite{boettcher2009patchy,hasegawa2010generating,boettcher2012ordinary,nogawa2014transition,singh2014scaling,boettcher2015classification}. 
Yakubo and Fujiki~\cite{yakubo2022general} discussed bond percolation on a general model of fractal SF networks using a renormalization approach, but their analysis was limited to locating the ordinary critical point $p_c$ via the percolating probability, rather than addressing the possibility of a critical phase (though, as demonstrated in this work, no critical phase is likely to emerge in finite-dimensional networks). 
Rozenfeld and Makse~\cite{rozenfeld2009fractality} reported simulation results for bond percolation on a loopy generalization of the SHM network covering both finite- and infinite-dimensional cases, but their analysis was limited to plotting the order parameter.
To our knowledge, no work has analytically examined the existence (or absence) of a critical phase in recursively constructed networks, explicitly contrasting finite-dimensional (fractal) and infinite-dimensional (small-world) regimes.

In this study, we formulate bond percolation on the SHM networks using a generating function approach and derive exact expressions for the average size of the root cluster (the cluster containing the node with the maximum degree) and for its fractal exponent. 
Using the fractal exponent of the root cluster, we show analytically and confirm numerically that the SHM network remains in a nonpercolating phase for all $p<1$ when the network is fractal (i.e., finite-dimensional), whereas it remains in a critical phase throughout the entire range $0<p<1$ when the network is small-world (i.e., infinite-dimensional), regardless of the degree exponent. 
These findings demonstrate that dimensionality---not degree heterogeneity---is a decisive factor for the appearance of a critical phase.

\section{Song--Havlin--Makse network}

\begin{figure}[t]
\centering
\includegraphics[width=12cm]{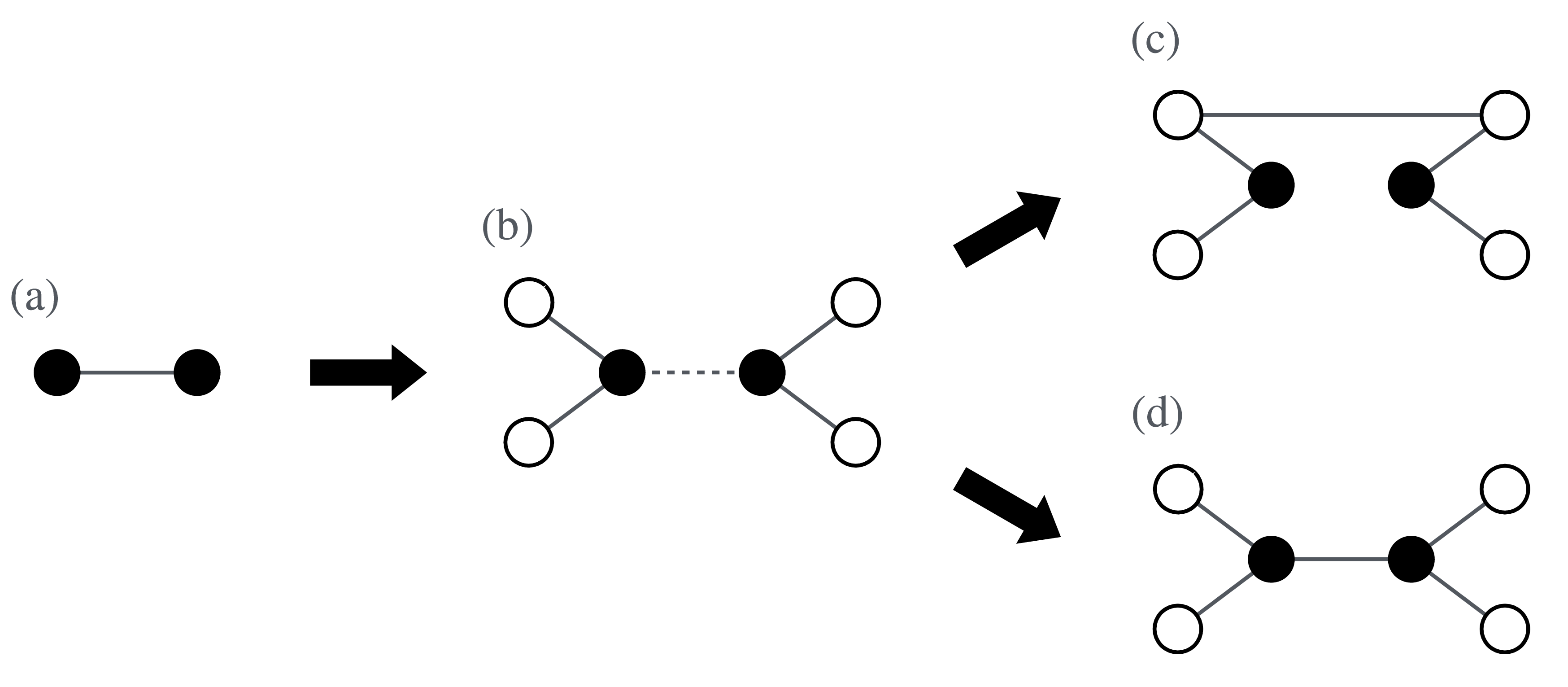}
\caption{
Illustration of the construction of the SHM network from $G_1$ to $G_2$, with $N_1=2$ and $m=2$.
(a) The initial network $G_1$ is a star graph with $N_1$ nodes.
(b) Each node $i$ (filled circles) in $G_1$ generates $mk_n(i)=2$ child nodes (open circles), which are connected to $i$.
(c) Each edge in $G_1$ (dotted lines in (b)) is rewired with probability $p_G$ by connecting a child of $i$ to a child of $j$. 
(d) Otherwise, the original edge is retained.
}
\label{fig:SHM-rewire}
\end{figure}

\begin{figure}[t]
\centering
(a)
\includegraphics[width=4.5cm]{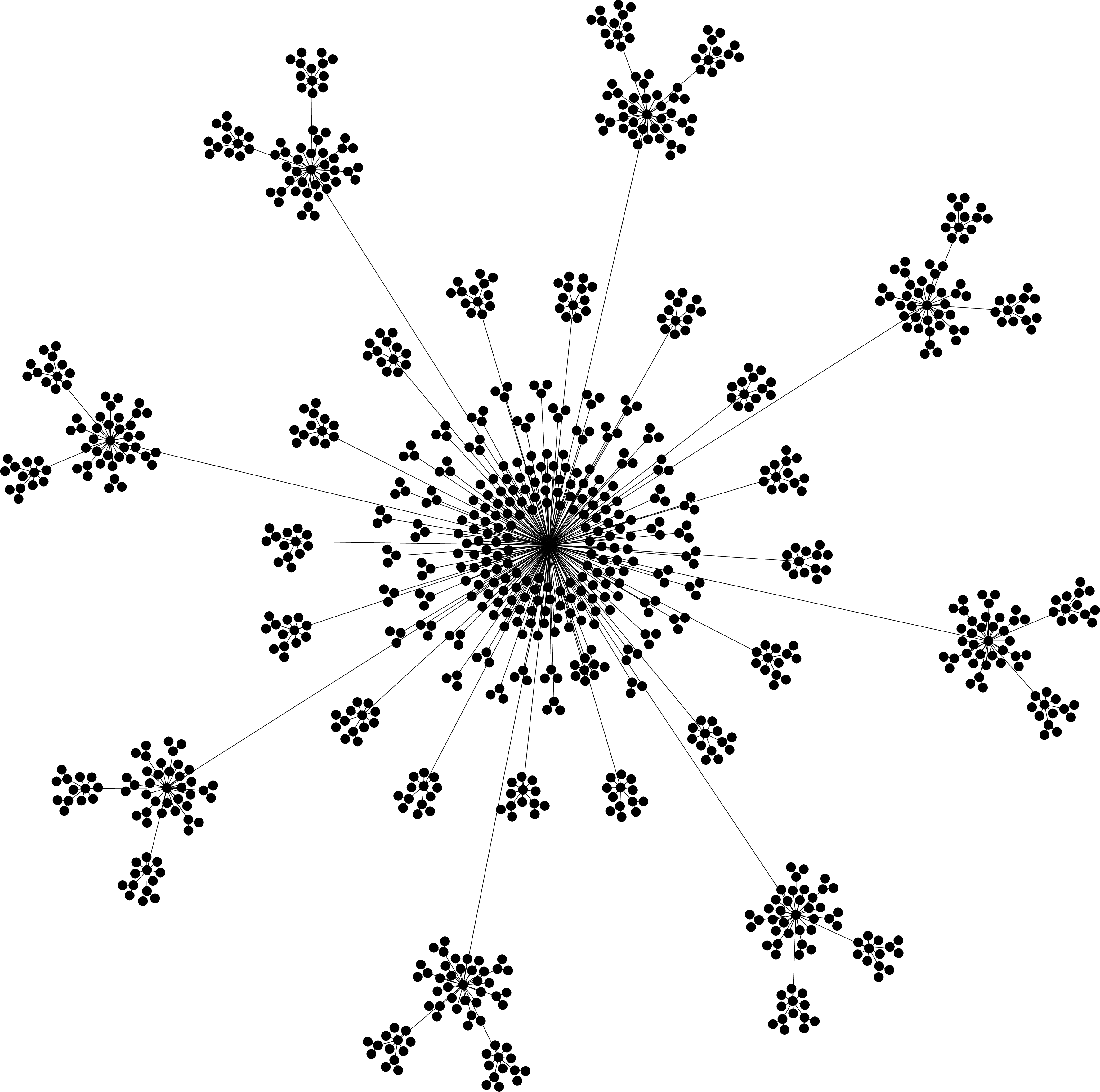}
(b)
\includegraphics[width=4.5cm]{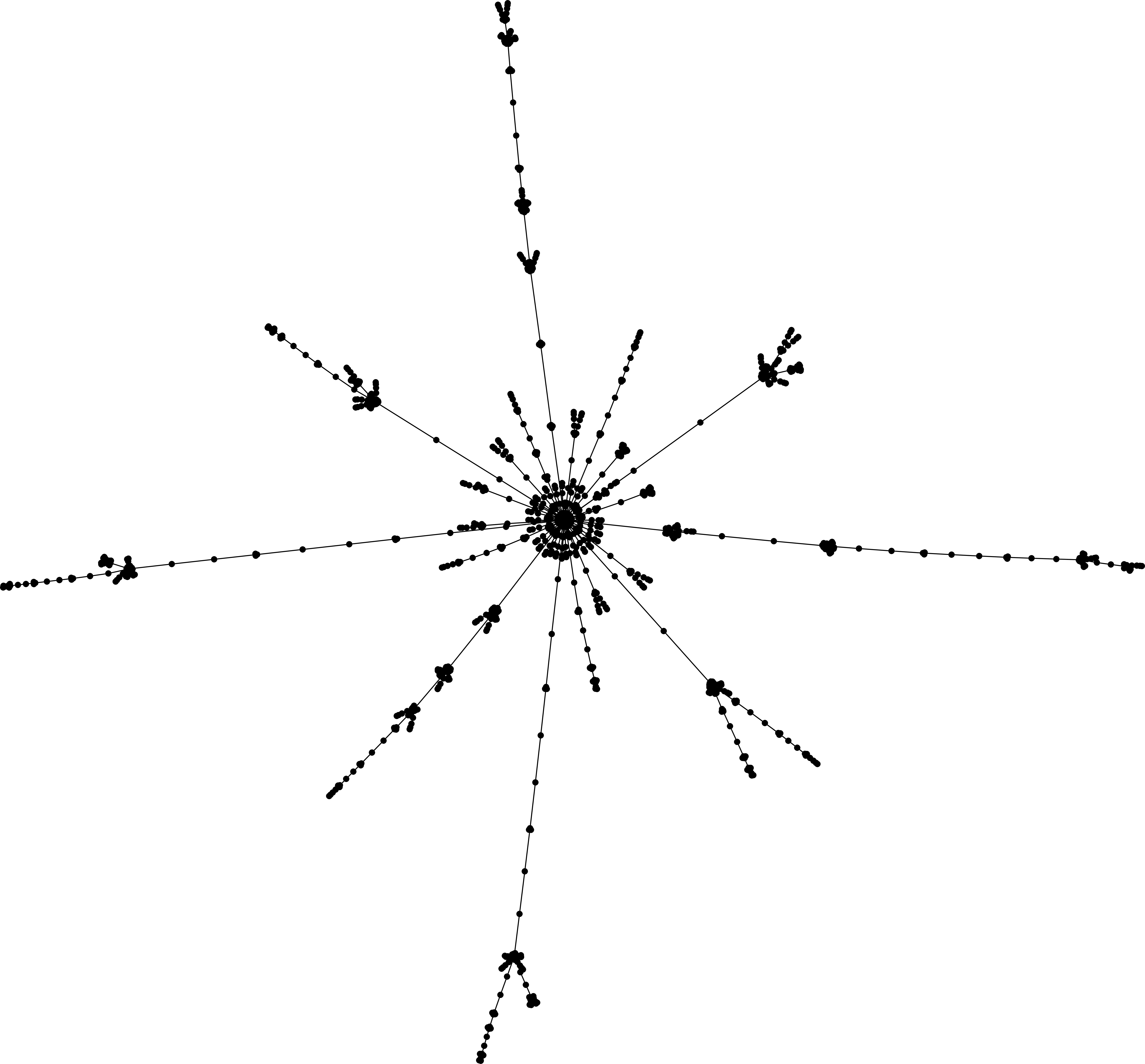}
(c)
\includegraphics[width=4.5cm]{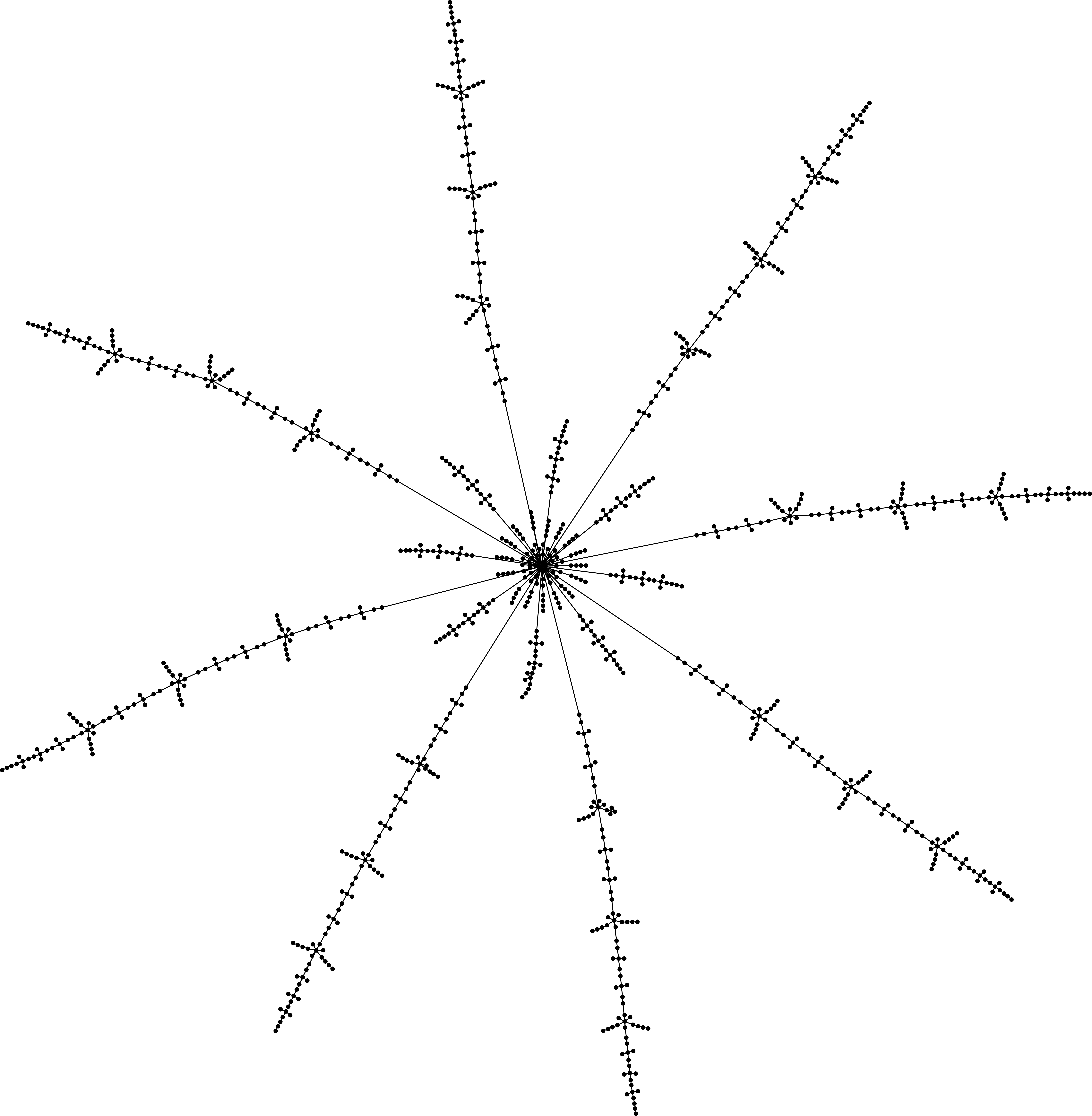}
\caption{
Examples of the SHM network $G_4$ generated with different rewiring probabilities: (a) $p_G=0$, (b) $p_G=0.5$, and (c) $p_G=1$.
In all cases, the initial number of nodes is $N_1=9$ and the branching parameter is $m=2$.
}
\label{fig:SHM-g4}
\end{figure}

This study focuses on the SHM networks~\cite{song2006origins}, denoted by $G_n$, where $n$ represents the generation.
We construct $G_n$ using a rewiring probability $p_G$ and a branching parameter $m$ ($\geq 2$) as follows.  
The initial network $G_1$ ($n = 1$) is a star graph with $N_1$ ($\geq 2$) nodes. 
Each subsequent network $G_{n+1}$ is recursively built from $G_n$ (see Fig.~\ref{fig:SHM-rewire}) according to the following rules: 
(i) For each node $i$ in $G_n$, add $m k_n(i)$ child nodes and connect them to node $i$, where $k_n(i)$ is the degree of node $i$ in $G_n$;
(ii) Rewire each edge $(i, j)$ in $G_n$ with probability $p_G$ by deleting $(i, j)$ and instead adding an edge between a child node of $i$ and a child node of $j$, both of which were added in step (i).
Figure~\ref{fig:SHM-g4} shows examples of $G_4$ for $p_G = 0$, $0.5$, and $1$.  
Any network constructed in this rule is a tree; it contains no cycles.

The structural properties of SHM networks have been studied in Refs.~\cite{song2006origins,yakubo2022general,yamamoto2023bifractality}.  
Let $N_n$ denote the number of nodes in $G_n$.  
By construction, $N_n$ is independent of the rewiring probability $p_G$ and is given by
\begin{equation}
N_n = (2m + 1)N_{n-1} - 2m = (2m + 1)^{n-1}(N_1 - 1) + 1.
\label{eq:Nn}
\end{equation}
Each node $i$ of degree $k_n(i)$ in $G_n$ is connected to $m k_n(i)$ new nodes in $G_{n+1}$, and each of its $k_n(i)$ existing edges is rewired with probability $p_G$.  
Hence, the expected degree of node $i$ in $G_{n+1}$ is $(m + 1 - p_G)k_n(i)$. 
The number of new nodes added in the construction of $G_{n_0+1}$ from $G_{n_0}$ is $2m(N_{n_0} - 1) = 2m(N_1 - 1)(2m + 1)^{n_0 - 1}$.
These added nodes have the average degree of $(m + 1 - p_G)^{n - n_0} (1+p_G/m)$ in $G_n$.
This recursive process generates a degree distribution $P(k)$ that follows a power law~\cite{song2006origins,yakubo2022general,yamamoto2023bifractality}:
\begin{equation}
P(k) \propto k^{-\gamma}.
\label{eq:degdis}
\end{equation}  
Here, the degree exponent $\gamma$ is given by  
\begin{equation}
\gamma = 1 + \frac{\log (2m + 1)}{\log (m + 1 - p_G)},  
\label{eq:m-expornent}
\end{equation}
which depends on both $p_G$ and $m$.
We denote the maximum degree in $G_n$ by $K_n$. 
Asymptotically, $K_n$ scales as
\begin{equation}
K_n \propto (m + 1 - p_G)^{n - 1} \quad (n \gg 1).  
\label{eq:Kn}
\end{equation}

While the SHM network remains SF regardless of $p_G$, its dimensionality depends strongly on $p_G$. 
Let $D_n$ denote the diameter of $G_n$, i.e. the longest shortest-path distance between any two nodes.  
When $p_G = 0$, meaning no edge rewiring occurs, the distance between two child nodes of parent nodes $i$ and $j$ is two steps longer than the distance between $i$ and $j$.  
Thus, the diameter $D_n$ evolves as
\begin{equation}
D_n = D_{n-1} + 2 = 2(n - 1) + D_1,  
\label{eq:sw-D}
\end{equation}  
where $D_1 = 1$ if $N_1 = 2$, and $D_1 = 2$ otherwise.  
Equation~(\ref{eq:sw-D}) implies that the average distance in $G_n$ scales at most as $O(n) = O(\log N_n)$, indicating that the network is a {\it small-world}~\cite{watts1998collective}, i.e, {\it infinite-dimensional}.
In contrast, for $0 < p_G \leq 1$, edges are rewired.  
If an edge $(i, j)$ is rewired, the shortest-path distance between $i$ and $j$ increases from 1 to 3: 
from $i$ to its child node, to a child node of $j$, and finally to $j$.  
In the deterministic case $p_G = 1$, all edges are rewired at each generation, and the diameter evolves as
\begin{equation}
D_n = 3 D_{n-1} + 2 = 3^{n - 1}(D_1 + 1) - 1.
\end{equation}  
For $0 < p_G < 1$, a fraction $p_G$ of edges are rewired, and asymptotically the diameter scales as~\cite{yamamoto2023bifractality}
\begin{equation}
D_n \propto (3 p_G + 1 - p_G)^{n - 1} = (1 + 2p_G)^{n - 1} \quad (n \gg 1).
\label{eq:m-D}
\end{equation}  
If the fractal dimension $d_{\mathrm{f}}$ is defined via the scaling relation $N_n \propto D_n^{d_{\mathrm{f}}}$, we obtain for $0 < p_G \leq 1$,
\begin{equation}
d_{\mathrm{f}} = \frac{\log (2m + 1)}{\log (1 + 2p_G)}.  
\label{eq:m-df}
\end{equation}  
In other words, for $0 < p_G \leq 1$, the SHM network is a {\it fractal} with a {\it finite} dimension that depends continuously on $p_G$.

In the next section, we focus on two deterministic cases: $p_G = 0$ and $p_G = 1$.  
Hereafter, we refer to the SHM network with $p_G = 0$ as the small-world SF tree, and that with $p_G = 1$ as the fractal SF tree.
We refer to the node(s) with maximum degree as the root(s), and define the root cluster as the cluster containing such a root. 
We focus on the root cluster, since the order of its size at a given $p$ indicates which phase the system belongs to.
We derive analytical expressions for the size of the root cluster, together with its associated fractal exponent, for the small-world SF tree and the fractal SF tree (the fractal exponent will be defined in the next section).

\section{Generating function approach}

\begin{figure}[t]
\centering
(a) 
\includegraphics[height=5cm]{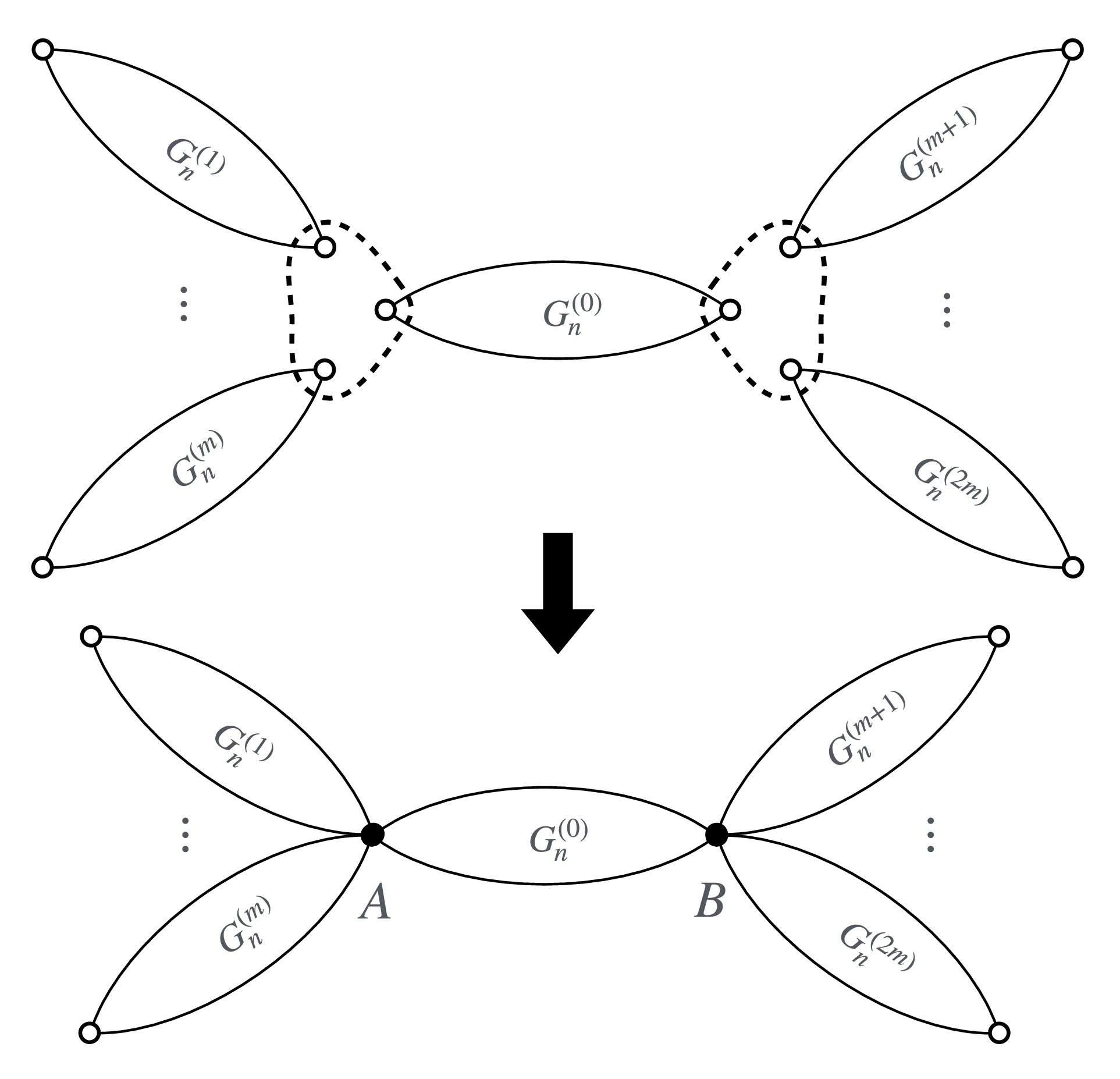}
(b) 
\includegraphics[height=5cm]{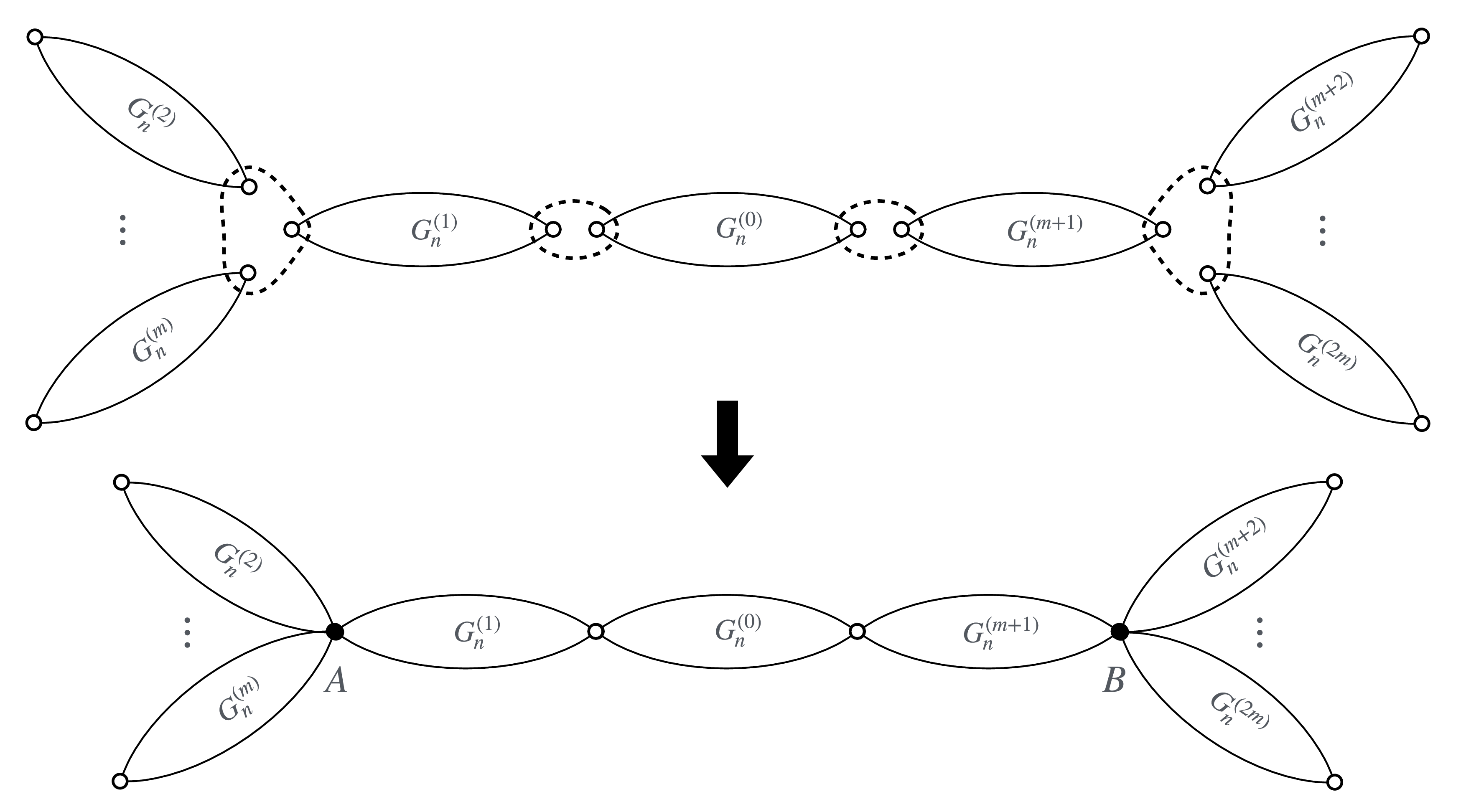}
\caption{
Recursive construction of $G_{n+1}$ from $2m+1$ copies of $G_n$ ($G_n^{(0)}, \cdots, G_n^{(2m)}$).
(a) For $p_G = 0$, $G_{n+1}$ is constructed by merging the roots of the $m$ copies $G_n^{(1)}, \ldots, G_n^{(m)}$ with root A of $G_n^{(0)}$, and the those of the remaining $m$ copies $G_n^{(m+1)}, \ldots, G_n^{(2m)}$ with root B of $G_n^{(0)}$.
(b) For $p_G = 1$, $G_{n+1}$ is constructed by merging the roots of $G_n^{(1)}$ and $G_n^{(m+1)}$ with the two roots of $G_n^{(0)}$, respectively. 
Then, the remaining root of $G_n^{(1)}$ is merged with the roots of the $m-1$ copies $G_n^{(2)}, \ldots, G_n^{(m)}$, and the remaining root of $G_n^{(m+1)}$ is merged with those of $G_n^{(m+2)}, \ldots, G_n^{(2m)}$. 
Open circles represent the roots of copies of $G_n$, while filled circles represent the two roots (A and B) of $G_{n+1}$.
}
\label{fig:config}
\end{figure}

In this section, we use a generating function approach to derive the average size of the root cluster and the fractal exponent for two limiting cases of the SHM network: the small-world SF tree ($p_G = 0$) and the fractal SF tree ($p_G = 1$).
We begin by focusing on the case $N_1 = 2$, which can be easily extended to the case $N_1 > 2$, as shown later.

As defined in the previous section, the root cluster is the cluster containing the root (the node with the maximum degree).  
When $N_1 = 2$, there are two equivalent roots (denoted A and B) in $G_n$; we focus on the root cluster containing one of them, namely root A.
To describe the root cluster, we introduce two probability distributions as follows. 
Let $w_n(s)$ denote the probability that root A is connected to root B in bond percolation on $G_n$ and belongs to a cluster of size $s$ (excluding the roots themselves).  
Similarly, let $v_n(s)$ denote the probability that root A is not connected to root B and belongs to a cluster of size $s$ (again, excluding the root).
The generating functions of $w_n(s)$ and $v_n(s)$ are defined as  
\begin{align}
W_n(x) &= \sum_{s=0}^\infty w_n(s) x^s, \\
\intertext{and}
V_n(x) &= \sum_{s=0}^\infty v_n(s) x^s,
\end{align}
respectively.
The generating function $W_n(x)$ encodes the size distribution of root clusters when the two roots are connected, while $V_n(x)$ corresponds to the case when they are disconnected.
We further define $P_n \equiv W_n(1) = \sum_s w_n(s)$ and $Q_n \equiv V_n(1) = \sum_s v_n(s)$, which represent the probabilities that the two roots in $G_n$ are connected and disconnected via retained edges, respectively.
By definition, the normalization condition $P_n + Q_n = W_n(1) + V_n(1) = 1$ holds for any $n$.

The generating function for the size distribution of the root cluster in $G_n$ is given in terms of $W_n(x)$ and $V_n(x)$ as $x^2 W_n(x) + x V_n(x)$. 
The factor $x^2$ accounts for the contribution of the two roots when the root cluster includes them both, while the factor $x$ corresponds to the contribution of root A when the root cluster contains only root A.
The average root cluster size $R_n(p)$ in $G_n$ is then given by
\begin{equation}
R_n(p)  = \left. \frac{d}{dx} \left( x^2 W_n(x) + x V_n(x) \right) \right|_{x=1} = 1 + P_n + W_n'(1) + V_n'(1),  
\label{eq:Rn-Gn}
\end{equation}
where $W_n'(1) = \sum_{s=0}^\infty s w_n(s)$ and $V_n'(1) = \sum_{s=0}^\infty s v_n(s)$.
Here, the first term on the right-hand side of Eq.~(\ref{eq:Rn-Gn}) corresponds to the contribution of root A, the second term corresponds to root B (when connected to root A), the third term corresponds to the contributions from other nodes in the root cluster containing both A and B, and the fourth term corresponds to the contribution from other nodes in the root cluster containing only root A.

Using the average root cluster size, we define the order parameter as  
\begin{equation}
\bar{s}_{\mathrm{root}}(p) = \frac{R_n(p)}{N_n}.
\end{equation}  
In the large size limit $n \to \infty$, the order parameter $\bar{s}_{\mathrm{root}}(p)$ is nonzero in the percolating phase and vanishes in both the nonpercolating and critical phases. 
To distinguish between the nonpercolating and critical phases, we introduce the {\it fractal exponent} $\psi(p)$~\cite{nogawa2009monte}.
By defining $\psi(p)$ via the scaling relation 
\begin{equation}
R_n(p) \propto N_n^{\psi(p)} \quad (n \gg 1),
\end{equation}
which quantifies how the root cluster scales with the system size, we characterize all three phases. 
The fractal exponent classifies them as follows:
(i) $\psi(p) = 0$ in the nonpercolating phase;  
(ii) $0 < \psi(p) < 1$ in the critical phase, where $\psi(p)$ varies continuously with $p$;  
(iii) $\psi(p) = 1$ in the percolating phase.
For computational convenience, we introduce the root cluster size and the number of nodes in $G_n$ excluding root A as $\tilde{R}_n(p) = R_n(p) - 1$ and $\tilde{N}_n = N_n - 1$, respectively.  
We then define a finite-size estimate of the fractal exponent for $G_n$ as
\begin{equation}
\psi_n(p) = \frac{\ln \tilde{R}_{n+1}(p) - \ln \tilde{R}_n(p)}{\ln \tilde{N}_{n+1} - \ln \tilde{N}_n}.  
\label{eq:psi-Gn}
\end{equation}  
In the large size limit, $\psi_n(p)$ approaches the fractal exponent:  
\begin{equation}
\psi(p) = \lim_{n \to \infty} \psi_n(p).
\end{equation}

\subsection{small-world SF tree}

The small-world SF tree $G_{n+1}$ is recursively constructed by combining identical copies of $G_n$ as follows (see Fig.~\ref{fig:config}(a)):  
(i) Prepare $2m + 1$ copies of $G_n$, denoted as $G_n^{(0)}, G_n^{(1)}, \ldots, G_n^{(2m)}$.  
(ii) Merge the roots on one side of the first $m$ copies, $G_n^{(1)}, \ldots, G_n^{(m)}$, with root A of $G_n^{(0)}$.
Merge the roots on one side of the remaining $m$ copies, $G_n^{(m+1)}, \ldots, G_n^{(2m)}$, with root B of $G_n^{(0)}$.
Here, `merge' means the identification of multiple nodes as a single node.
(iii) These two merged nodes are then assigned as root A and B of $G_{n+1}$, respectively.

From the construction shown in Fig.~\ref{fig:config}(a), we obtain the following recurrence relations for the generating functions $W_n(x)$ and $V_n(x)$ in the small-world SF tree:  
\begin{align}
W_{n+1}(x) &= W_n(x) [x W_n(x) + V_n(x)]^{2m},  
\label{eq:Wn-smallworld} \\
V_{n+1}(x) &= V_n(x) [x W_n(x) + V_n(x)]^m,  
\label{eq:Vn-smallworld}
\end{align}  
where the initial condition is $W_1(x) = p$ and $V_1(x) = 1 - p$.
In Eq.~(\ref{eq:Wn-smallworld}), the factor $W_n(x)$ represents the contribution from $G_n^{(0)}$ when its roots A and B are connected.  
The term $(x W_n(x) + V_n(x))^{2m}$ accounts for the contributions from the $2m$ copies $G_n^{(1)}, \ldots, G_n^{(2m)}$, which are connected to root A or root B of $G_n^{(0)}$ via the merged roots.
Equation~(\ref{eq:Vn-smallworld}) corresponds to the case where root A and B of $G_n^{(0)}$ are not connected.  
The factor $V_n(x)$ represents the contribution from $G_n^{(0)}$ in this case, and $(x W_n(x) + V_n(x))^m$ accounts for the contributions from the $m$ copies $G_n^{(1)}, \ldots, G_n^{(m)}$ connected to root A.
From Eqs.~(\ref{eq:Wn-smallworld}) and~(\ref{eq:Vn-smallworld}), it follows that  
\begin{equation}
P_n = p \quad \text{and} \quad Q_n = 1 - p,  
\label{eq:Pn-smallworld}
\end{equation}  
for all $n$.  
This result reflects the fact that the edge connecting root A and  B (i.e., the edge that constitutes $G_1$) is never rewired when $p_G = 0$, and that a tree contains only one unique path between any pair of nodes.  
Hence, the probability that the two roots in $G_n$ are connected remains $p$, regardless of $n$.

From Eqs.~(\ref{eq:Wn-smallworld}) and~(\ref{eq:Vn-smallworld}), we derive recurrence relations for the derivatives $W_n'(1)$ and $V_n'(1)$:  
 \begin{align}
W_{n+1}'(1) &= W_n'(1) + 2m P_n \left[ P_n + W_n'(1) + V_n'(1) \right], \\
V_{n+1}'(1) &= V_n'(1) + m Q_n \left[ P_n + W_n'(1) + V_n'(1) \right].
\end{align}
Summing these equations yields a recurrence relation for $W_{n+1}'(1) + V_{n+1}'(1)$:  
\begin{equation}
W_{n+1}'(1) + V_{n+1}'(1) = (1 + m + m P_n) \left[ P_n + W_n'(1) + V_n'(1) \right] - P_n.  
\label{eq:RecEq-smallworld}
\end{equation}
Using Eq.~(\ref{eq:Pn-smallworld}) and Eq.~(\ref{eq:RecEq-smallworld}), we obtain $\tilde{R}_n(p) = P_n + W_n'(1) + V_n'(1)$, i.e., the root cluster size excluding root A, as 
\begin{equation}
\tilde{R}_n(p) 
= (1 + m + m p)\, \tilde{R}_{n-1}(p) 
= (1 + m + m p)^2\, \tilde{R}_{n-2}(p)
= \cdots = p (1 + m + m p)^{n-1},  
\label{eq:Rn-smallworld}
\end{equation}
where we used the initial condition $\tilde{R}_1(p) = P_1 + W_1'(1) + V_1'(1) = p$.
Thus the average root cluster size $R_n(p)$ of $G_n$ is given by
\begin{equation}
R_n(p) = 1 + p (1 + m + m p)^{n-1}.  
\label{eq:root-smallworld}
\end{equation}
Substituting Eq.~(\ref{eq:Rn-smallworld}) into Eq.~(\ref{eq:psi-Gn}), we obtain the fractal exponent $\psi_n(p)$ for $G_n$:
\begin{equation}
\psi_n(p) 
= \frac{ \ln \left[ p (1 + m + m p)^n \right] - \ln \left[ p (1 + m + m p)^{n-1} \right] }{ \ln (1 + 2m)^n - \ln (1 + 2m)^{n-1} } 
= \frac{ \ln (1 + m + m p) }{ \ln (1 + 2m) }.
\end{equation}
Taking the limit $n \to \infty$, we obtain the fractal exponent for the small-world SF tree:  
\begin{equation}
\psi(p) = \lim_{n \to \infty} \psi_n(p) = \frac{ \ln (1 + m + m p) }{ \ln (1 + 2m) }.  
\label{eq:psi-smallworld}
\end{equation}

Furthermore, the generating function formalism enables the evaluation of the cluster size distribution $n(s)$, i.e., the average number of clusters of size $s$ per node.   
Let $u_n(s)$ denote the average number of clusters of size $s$ in bond percolation (with probability $p$) on $G_n$, excluding the cluster(s) containing the root(s).   
The corresponding generating function is defined as 
\begin{equation}
U_n(x) = \sum_{s=0}^\infty u_n(s) x^s.
\end{equation}
Based on the construction of $G_{n+1}$ shown in Fig.~\ref{fig:config}(a), $U_{n+1}(x)$ can be given in terms of $U_n(x)$ and $V_n(x)$ as follows:  
\begin{equation}
U_{n+1}(x) = (2m + 1) U_n(x) + 2m x V_n(x).  
\label{eq:Un-smallworld}
\end{equation}  
The first term on the right-hand side represents the contribution from clusters not connected to the roots in each of the $2m + 1$ copies of $G_n$.  
The second term corresponds to the root clusters in the $2m$ copies $G_n^{(1)}, \ldots, G_n^{(2m)}$ that are not connected to either root A or B of $G_{n+1}$.
We define $\tilde{U}_n(x) \equiv \sum_{s=0}^\infty n(s) x^s$ as the generating function for the cluster size distribution $n(s)$ in $G_n$.
Since clusters both containing and not containing the roots are counted in $n(s)$,
\begin{equation}
\tilde{U}_n(x) = U_n(x) + 2 x V_n(x) + x^2 W_n(x).
\label{eq:tildeUn}
\end{equation}
Given $\tilde{U}_n(x)$, we evaluate the cluster size distribution $n(s)$ through  
\begin{equation}
n(s) = \frac{1}{N_n} \cdot \frac{1}{s!} \left. \frac{d^s}{dx^s} \tilde{U}_n(x) \right|_{x=0}.  
\label{eq:ns}
\end{equation}

The case $N_1 > 2$ can be analyzed using the same formalism as above.  
Let $G_{n, N_1}$ denote the small-world SF tree of generation $n$ constructed from an initial network with $N_1 > 2$ nodes.  
It is easy to see that the network $G_{n, N_1}$ is constructed by merging the roots of $N_1 - 1$ copies of $G_n$ (each constructed with $N_1 = 2$) into a single root, which becomes the root of $G_{n, N_1}$.
Since the $N_1 - 1$ subtrees connected to the root are independent of one another, the generating function for the root cluster size of $G_{n, N_1}$ is given by $x \left( x W_n(x) + V_n(x) \right)^{N_1 - 1}$, where $W_n(x)$ and $V_n(x)$ satisfy Eqs.~(\ref{eq:Wn-smallworld}) and~(\ref{eq:Vn-smallworld}), respectively.
The average root cluster size $R_{n, N_1}(p)$ is then given by
\begin{equation}
R_{n, N_1}(p) = 1 + (N_1 - 1)\, \tilde{R}_n(p),  
\label{eq:Rn-general}
\end{equation}  
where $\tilde{R}_n(p)$ is given by Eq.~(\ref{eq:Rn-smallworld}) for $G_n$.
It immediately follows from Eq.~(\ref{eq:Rn-general}) that the fractal exponent $\psi(p)$ is independent of $N_1$.   
In other words, for small-world SF trees with arbitrary $N_1 > 2$, the fractal exponent $\psi(p)$ remains given by Eq.~(\ref{eq:psi-smallworld}).
Furthermore, the generating function $\tilde{U}_{n, N_1}(x)$ for the cluster size distribution of $G_{n, N_1}$ is  
\begin{equation}
\tilde{U}_{n, N_1}(x) = (N_1 - 1)\left( U_n(x) + x V_n(x) \right) + x \left( x W_n(x) + V_n(x) \right)^{N_1 - 1},  
\label{eq:Un-general}
\end{equation}  
where $U_n(x)$ is given by Eq.~(\ref{eq:Un-smallworld}).

\subsection{fractal SF tree}

In the case of the fractal SF tree ($p_G = 1$), $G_{n+1}$ can be constructed recursively from copies of $G_n$ (see Fig.~\ref{fig:config}(b)):  
(i) Prepare $2m + 1$ copies of $G_n$, denoted as $G_n^{(0)}, G_n^{(1)}, \ldots, G_n^{(2m)}$.  
(ii) Merge the two roots of $G_n^{(0)}$ with the roots of $G_n^{(1)}$ and $G_n^{(m+1)}$, respectively.  
(iii) Merge the remaining root of $G_n^{(1)}$ with the roots on one side of the $m - 1$ copies $G_n^{(2)}, \ldots, G_n^{(m)}$.
Likewise, merge the remaining root of $G_n^{(m+1)}$ with the roots on one side of the $m - 1$ copies $G_n^{(m+2)}, \ldots, G_n^{(2m)}$.  
(iv) Assign the two merged nodes obtained in step (iii) as roots A and B of $G_{n+1}$.

Based on the construction shown in Fig.~\ref{fig:config}(b), the generating functions $W_n(x)$ and $V_n(x)$ satisfy the following recurrence relations:  
\begin{align}
W_{n+1}(x) &= x^2 W_n^3(x) [x W_n(x) + V_n(x)]^{2m - 2},  
\label{eq:Wn-fractal} \\
V_{n+1}(x) &= \left[ x^2 W_n^2(x) V_n(x) + x W_n(x) V_n(x) + V_n(x) \right] [x W_n(x) + V_n(x)]^{m - 1},  
\label{eq:Vn-fractal}
\end{align}
where the initial condition is $W_1(x) = p$ and $V_1(x) = 1 - p$.
Root A and B in $G_{n+1}$ are connected only via the three copies $G_n^{(0)}$, $G_n^{(1)}$, and $G_n^{(m+1)}$, which give rise to the term $x^2 W_n^3(x)$ in Eq.~(\ref{eq:Wn-fractal}). 
The factor $x^2$ represents the contribution from the two merged roots on $G_n^{(0)}$.  
The term $(x W_n(x) + V_n(x))^{2m - 2}$ corresponds to the contribution from the remaining $2m - 2$ copies of $G_n$---namely, $G_n^{(2)}, \ldots, G_n^{(m)}$ and $G_n^{(m+2)}, \ldots, G_n^{(2m)}$.
The term $x^2 W_n^2(x) V_n(x) + x W_n(x) V_n(x) + V_n(x)$ in Eq.~(\ref{eq:Vn-fractal}) accounts for the combined contributions involving $G_n^{(1)}$, $G_n^{(0)}$, and $G_n^{(m+1)}$ when the root cluster does not include root B.
The term $(x W_n(x) + V_n(x))^{m - 1}$ accounts for the contributions from the $m - 1$ copies ($G_n^{(2)}, \ldots, G_n^{(m)}$) connected to root A.
From Eq.~(\ref{eq:Wn-fractal}), we find that the probability $P_n$ that root A and B belong to the same cluster satisfies the recurrence relation  
\begin{equation}
P_{n+1} = P_n^3, \label{eq:P-fractal}
\end{equation}  
which leads to  
\begin{equation}
P_n = p^{3^{n - 1}} \quad \text{and} \quad Q_n = 1 - p^{3^{n - 1}}.
\end{equation}

For the derivatives $W_n'(1)$ and $V_n'(1)$, we obtain the following recurrence relations from Eqs.~(\ref{eq:Wn-fractal}) and~(\ref{eq:Vn-fractal}):  
\begin{align}
W_{n+1}'(1) &= 2 P_n^3 + 3 P_n^2 W_n'(1) + 2(m - 1) P_n^3 \left[ P_n + W_n'(1) + V_n'(1) \right], \\
V_{n+1}'(1) &= (1 - P_n^3)(m - 1)\left[ P_n + W_n'(1) + V_n'(1) \right] \nonumber \\
&\quad + (P_n + 2 P_n^2) Q_n + (1 + 2 P_n) Q_n W_n'(1) + (1 + P_n + P_n^2) V_n'(1).
\end{align}
By combining these, we obtain the recurrence relation for $W_n'(1) + V_n'(1)$ as
\begin{align}
W_{n+1}'(1) + V_{n+1}'(1) 
= \left[ m + P_n + P_n^2 + (m - 1) P_n^3 \right] \left[ P_n + W_n'(1) + V_n'(1) \right] - P_n^3.  
\label{eq:f-rsABandrsA}
\end{align}
Using Eq.~(\ref{eq:P-fractal}) and the initial condition $\tilde{R}_1(p) = p$, we obtain $\tilde{R}_n(p)$ as
\begin{equation}
\tilde{R}_n(p) 
= \left[ m + P_{n-1} + P_{n-1}^2 + (m - 1) P_{n-1}^3 \right] \tilde{R}_{n-1}(p)
= \cdots
= p \prod_{n' = 1}^{n - 1} \left[ m + p^{3^{n' - 1}} + p^{2 \cdot 3^{n' - 1}} + (m - 1) p^{3^{n'}} \right],  
\label{eq:Rn-fractal}
\end{equation}
for $n \geq 2$.
Therefore, the average root cluster size $R_n(p)$ of $G_n$ is, for $n \geq 2$,
\begin{equation}
R_n(p) = 1 + p \prod_{n' = 1}^{n - 1} \left[ m + p^{3^{n' - 1}} + p^{2 \cdot 3^{n' - 1}} + (m - 1) p^{3^{n'}} \right],  
\label{eq:root-fractal}
\end{equation}
and $R_1(p) = 1 + p$.
Substituting Eq.~(\ref{eq:Rn-fractal}) into Eq.~(\ref{eq:psi-Gn}), we obtain the fractal exponent $\psi_n(p)$ for $G_n$ as
\begin{equation}
\psi_n(p) = \frac{ \ln \left[ m + p^{3^{n - 1}} + p^{2 \cdot 3^{n - 1}} + (m - 1) p^{3^n} \right] }{ \ln (1 + 2m) }.
\end{equation}
Taking the limit $n \to \infty$, we obtain the fractal exponent $\psi(p)$ for $p < 1$ as
\begin{equation}
\psi(p) = \lim_{n \to \infty} \psi_n(p) = \frac{ \log m }{ \log (2m + 1) }.  
\label{eq:psi-fractal}
\end{equation}

We now consider the generating function $U_n(x)$ for the cluster size distribution $u_n(s)$, excluding any clusters that contain either of the roots. 
According to the construction of $G_{n+1}$ shown in Fig.~\ref{fig:config}(b), $U_{n+1}(x)$ can be given in terms of $U_n(x)$, $W_n(x)$, and $V_n(x)$ as follows:  
\begin{equation}
U_{n+1}(x) = (2m + 1) U_n(x) + 2x V_n^2(x) + x^2 W_n(x) V_n^2(x) + (2m - 2)x V_n(x).  
\label{eq:Un-fractal}
\end{equation}
The first term on the right-hand side represents the contribution from clusters not connected to any roots in the $2m + 1$ copies of $G_n$.  
The second and third terms represent contributions from clusters involving the merged roots between $G_n^{(0)}$ and $G_n^{(1)}$, and between $G_n^{(0)}$ and $G_n^{(m+1)}$.  
The second term corresponds to the case where the two merged roots are not connected, while the third term corresponds to the case where they are connected.  
The fourth term accounts for root clusters in the remaining $2m - 2$ copies ($G_n^{(2)}, \ldots, G_n^{(m)}$ and $G_n^{(m+2)}, \ldots, G_n^{(2m)}$) that are not connected to either of the roots A or B in $G_{n+1}$.  
With $W_n(x)$ and $V_n(x)$ obtained from Eqs.~(\ref{eq:Wn-fractal}) and~(\ref{eq:Vn-fractal}), and $U_n(x)$ from Eq.~(\ref{eq:Un-fractal}), we compute $\tilde{U}_n(x)$ via Eq.~(\ref{eq:tildeUn}).  
Substituting this into Eq.~(\ref{eq:ns}), we numerically compute the cluster size distribution $n(s)$.

As in the case of small-world SF trees, the root cluster size and cluster size distribution for $N_1 > 2$ can be readily determined from the results for $N_1 = 2$.  
The generating function for the root cluster size of $G_{n, N_1}$, constructed from an initial network with $N_1$ nodes, is given by  $x \left( x W_n(x) + V_n(x) \right)^{N_1 - 1}$, where $W_n(x)$ and $V_n(x)$ are obtained from Eqs.~(\ref{eq:Wn-fractal}) and~(\ref{eq:Vn-fractal}), respectively, for the case $N_1 = 2$.
The corresponding average root cluster size $R_{n, N_1}(p)$ is given by substituting Eq.~(\ref{eq:Rn-fractal}) into Eq.~(\ref{eq:Rn-general}).
It follows that the fractal exponent remains given by Eq.~(\ref{eq:psi-fractal}).
Similarly, the generating function $\tilde{U}_{n, N_1}(x)$ for the cluster size distribution of $G_{n, N_1}$ is given by Eq.~(\ref{eq:Un-general}), where $U_n(x)$ is taken from Eq.~(\ref{eq:Un-fractal}) for $G_n$.

\section{Percolation properties of the SHM network}

\begin{figure}[t]
\centering
(a) \includegraphics[height=5.5cm]{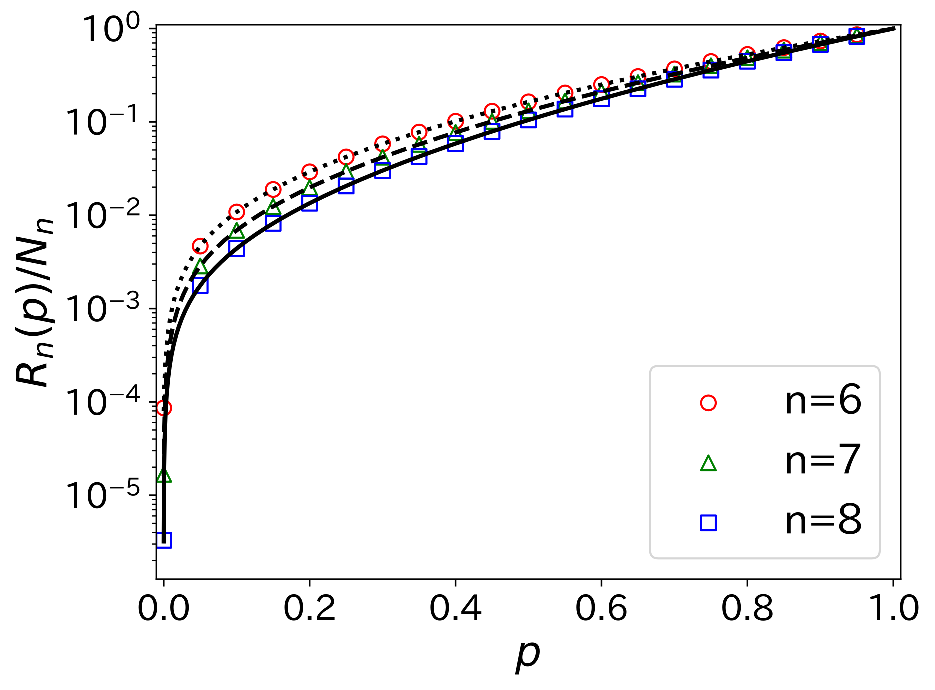}
(b) \includegraphics[height=5.5cm]{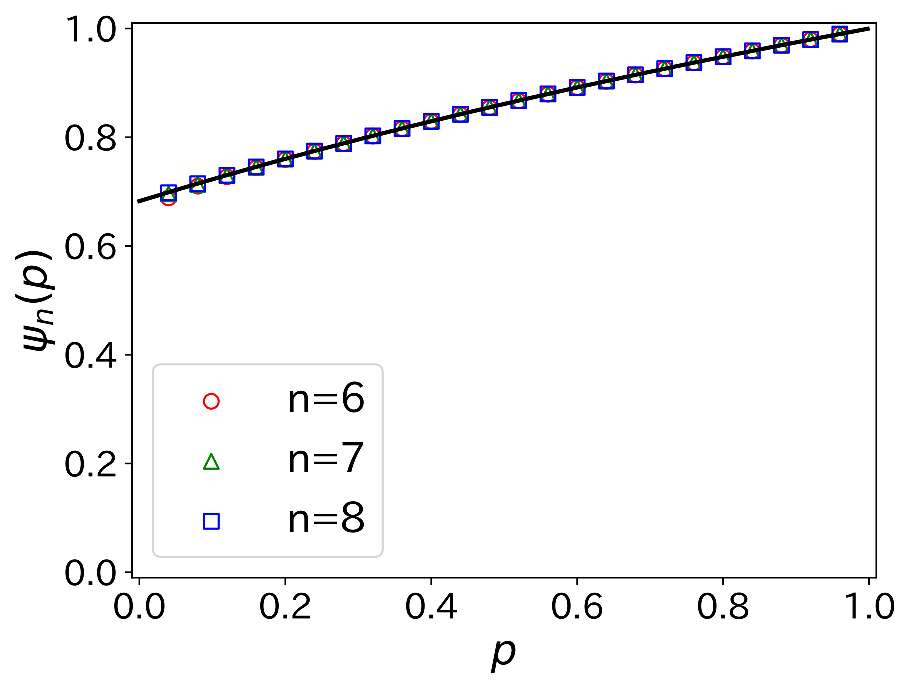}
\caption{
(a) $\bar{s}_{\mathrm{root}} (p) = R_n(p)/N_n$ and (b) $\psi_n(p)$ as a function of $p$ in the small-world SF tree.
Red circles, green triangles, and blue squares represent Monte Carlo results for $G_n$ with $n=6,7$ and $8$, respectively.
In panel (a), the black dotted, dashed, and solid lines represent the theoretical predictions of $R_n(p)/N_n$ for $G_6$, $G_7$, and $G_8$, respectively, obtained from Eq.~(\ref{eq:Nn}) and Eq.~(\ref{eq:root-smallworld}).
In panel (b), the black line represents the theoretical curve of the fractal exponent $\psi(p)$, obtained from Eq.~(\ref{eq:psi-smallworld}).
}\label{fig:mANDpsi-smallworld}
\end{figure}

In this section, we examine the percolation properties of the SHM networks.  
To validate the generating function analysis presented in the previous section, we performed Monte Carlo simulations of bond percolation on the SHM networks. 
For each value of $p_G$, we generated independent 2500 network realizations and applied the Newman--Ziff algorithm~\cite{newman2001fast}, conducting 200 independent runs for each network.   
Throughout this section, we fix the number of nodes in $G_1$ to  $N_1 = 5$.  
As shown in all figures presented below, the results from the generating function approach are in excellent agreement with those from Monte Carlo simulations for both $p_G = 0$ and $p_G = 1$.

We begin by focusing on the small-world SF tree ($p_G = 0$), which is infinite-dimensional.  
Figure~\ref{fig:mANDpsi-smallworld}(a) shows the order parameter $\bar{s}_{\mathrm{root}}(p)$ as a function of $p$ for generations $n = 6$, $7$, and $8$.  
The order parameter $\bar{s}_{\mathrm{root}}(p)$ decreases with increasing $n$ and tends toward zero for all values of $p$.
According to Eq.~(\ref{eq:root-smallworld}), $\bar{s}_{\mathrm{root}}(p)$ vanishes in the limit $n \to \infty$ for all $p < 1$, indicating the absence of the percolating phase; that is, $p_{c2} = 1$.
Figure~\ref{fig:mANDpsi-smallworld}(b) presents the fractal exponent $\psi_n(p)$ of $G_n$ as a function of $p$.  
The Monte Carlo results for $\psi_n(p)$ converge to the theoretical curve~(\ref{eq:psi-smallworld}) for $\psi(p)$ as $n$ increases.  
Equation~(\ref{eq:psi-smallworld}) shows that the fractal exponent $\psi(p)$ varies continuously for $0 < p < 1$, increasing from $\psi = \log 3 / \log 5$ at $p \approx 0$ to $\psi = 1$ at $p \approx 1$.
These results indicate that bond percolation on the small-world SF tree remains in the critical phase throughout the entire range of $p$, i.e., $p_{c1}=0$ and $p_{c2}=1$.

\begin{figure}[t]
\centering
(a) \includegraphics[height=3.8cm]{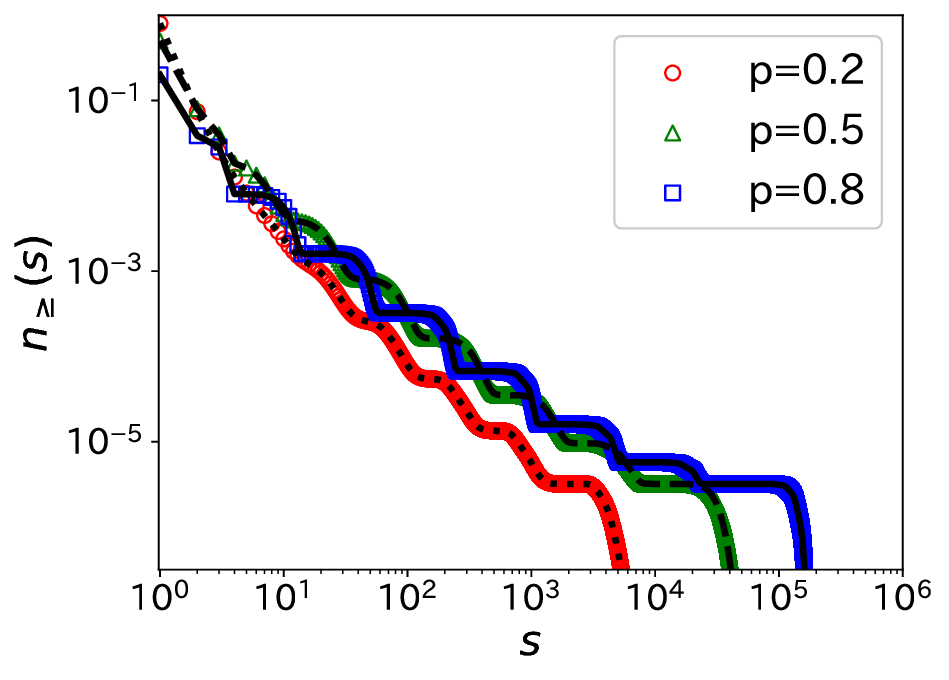}
(b) \includegraphics[height=3.8cm]{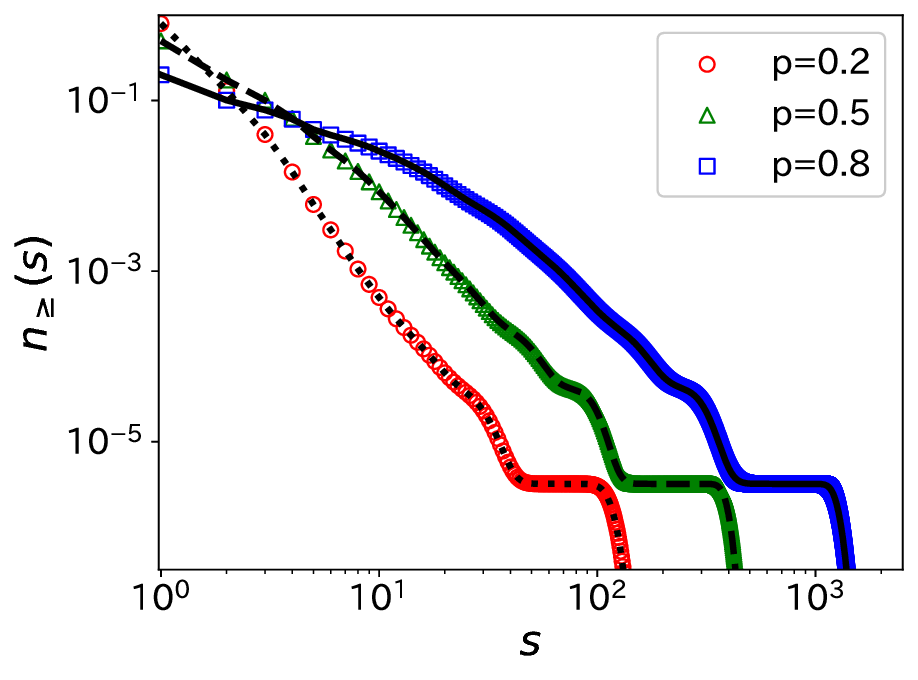}
(c) \includegraphics[height=3.8cm]{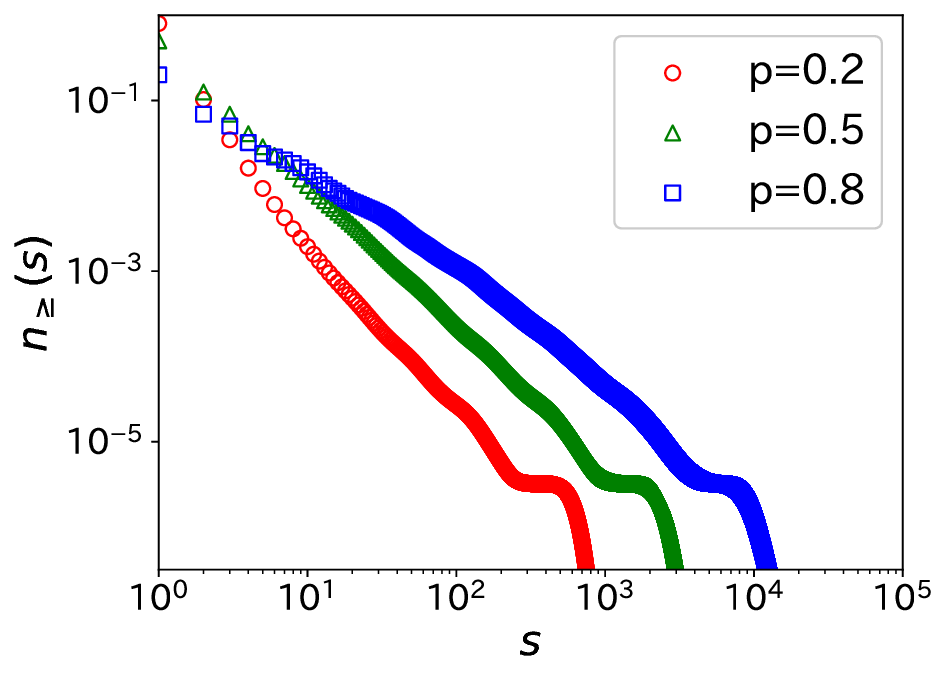}
\caption{
Cumulative cluster size distribution $n_{\ge}(s) = \sum_{s^\prime \geq s} n(s^\prime)$ for bond percolation with $p=0.2$, $0.5$, and $0.8$ on (a) small-world SF tree, (b) fractal SF tree, and (c) the SHM network with $p_G=0.5$, all at generation $n=8$ ($N_n = 312,501$).
Red circles, green triangles, and blue squares represent Monte Carlo simulation results for $p=0.2$, $0.5$, and $0.8$, respectively.
In panels (a) and (b), the black dotted, dashed, and solid lines represent numerical evaluations obtained from the generating function method.
}\label{fig:ns}
\end{figure}

\begin{figure}[t]
\centering
(a) \includegraphics[height=4cm]{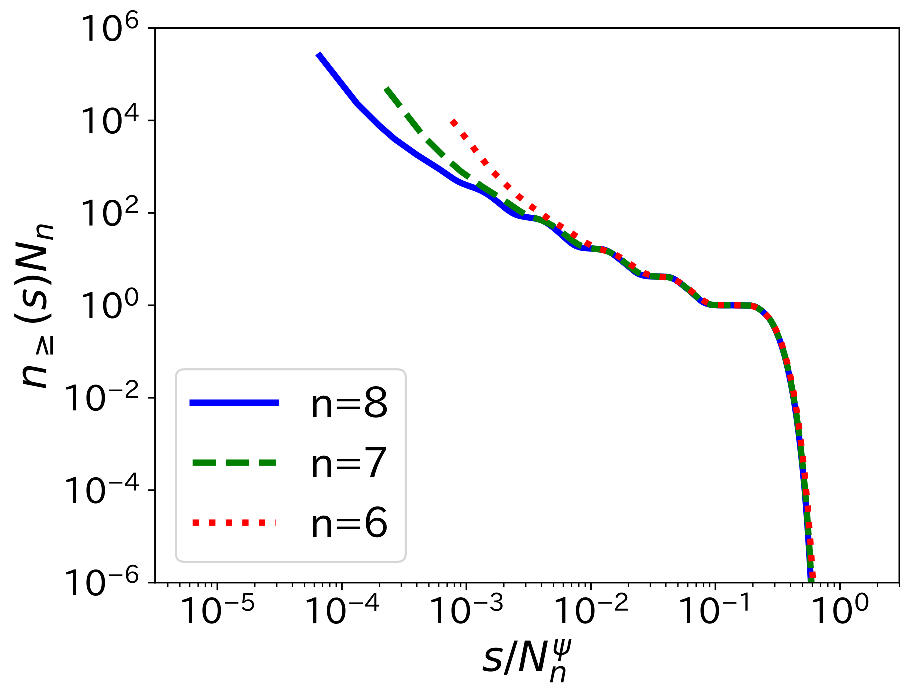}
(b) \includegraphics[height=4cm]{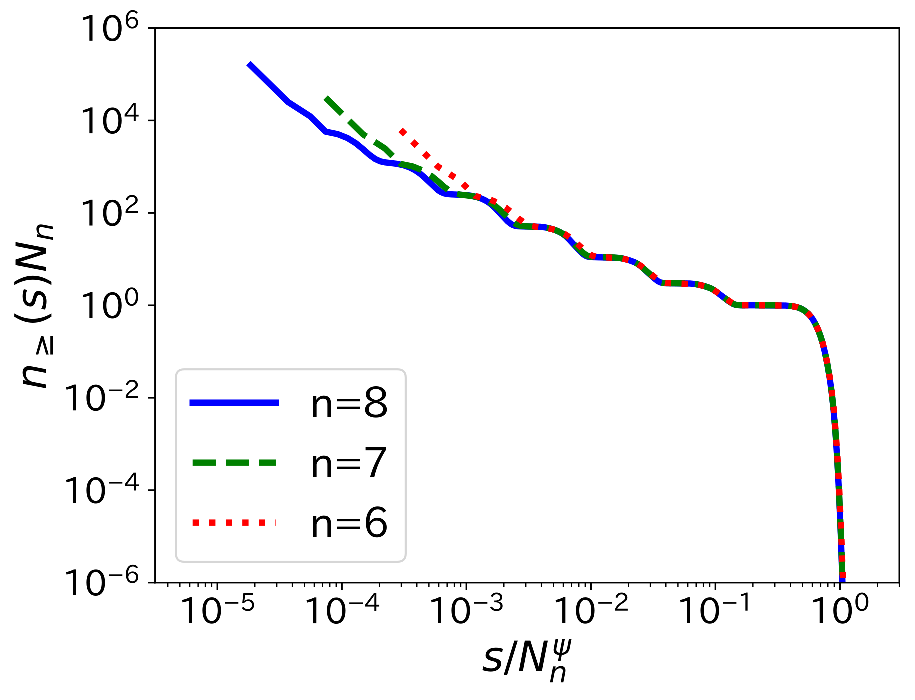}
(c) \includegraphics[height=4cm]{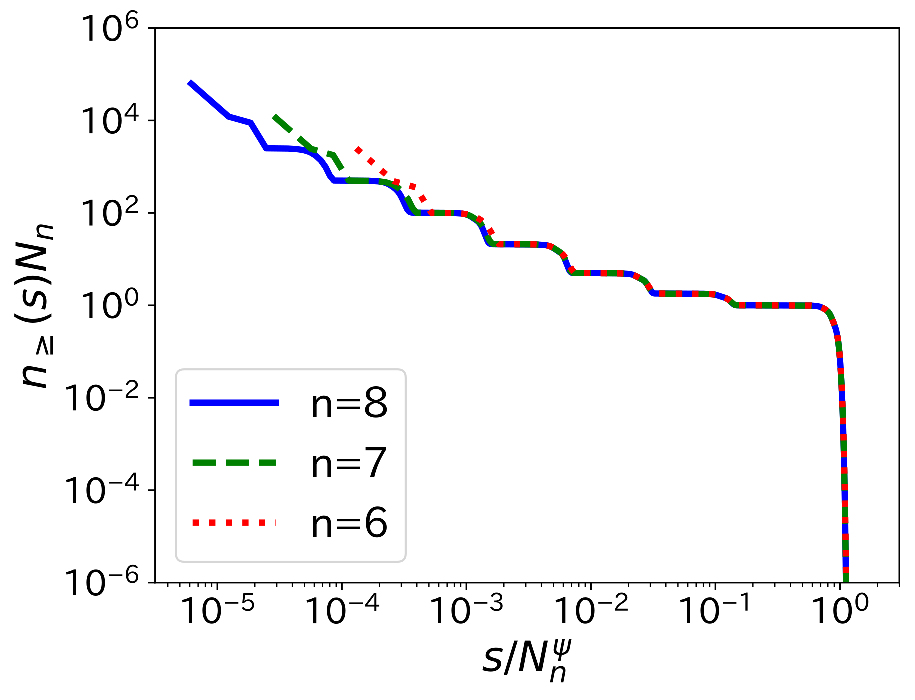}
\caption{
Finite-size scaling of the cumulative cluster size distribution $n_{\geq}(s)$, rescaled as $s/N_n^{\psi_n}$ versus $n_{\geq}(s)N_n$, for the small-world SF tree at (a) $p = 0.2$, (b) $p = 0.5$, and (c) $p = 0.8$. 
Curves are calculated using the generating functions (Eq.~(\ref{eq:ns})) for $G_6$ (red dotted), $G_7$ (green dashed), and $G_8$ (blue solid).
}\label{fig:sw-nsp}
\end{figure}

Figure~\ref{fig:ns}(a) shows the cumulative cluster size distribution $n_{\geq}(s) = \sum_{s^\prime \geq s} n(s^\prime)$ for $p = 0.2$, $0.5$, and $0.8$, obtained from both Monte Carlo simulations and the generating function approach.
The results demonstrate that the cumulative distributions computed using the generating functions are in good agreement with those obtained from the Monte Carlo simulations.

We carry out finite-size scaling of the cumulative cluster size distribution $n_{\geq}(s)$ for several values of $p$, and confirm that it follows a power-law form, with the exponent varying with $p$.  
We assume the following finite-size scaling form for $n_{\geq}(s)$:
\begin{equation}
n_{\geq}(s) = N_n^{-\psi(p)(\tau(p) - 1)} f\left(s N_n^{-\psi(p)}\right),
\end{equation}  
where scaling function $f(x)$ behaves as  
\begin{equation}
f(x) \sim 
\begin{cases}
\text{rapidly decaying function} & \text{for} \quad x \gg 1, \\
x^{1 - \tau(p)} & \text{for} \quad x \ll 1.
\end{cases}
\end{equation}
We further assume that the exponent $\tau(p)$ is related to the fractal exponent $\psi(p)$ via
\begin{equation}
\tau(p) = 1 + \psi^{-1}(p), \label{eq:tau} 
\end{equation}  
as reported in Ref.~\cite{nogawa2009monte}.
We remark that the scaling relation (\ref{eq:tau}) can be regarded as a natural generalization of that for Euclidean lattices. 
In Euclidean lattices, where a linear dimension is defined, one has $\psi(p_c) =  \tilde{d}/d$ ($d$ is the lattice dimension and $\tilde{d}$ is the fractal dimension of the largest cluster), so that Eq.~(\ref{eq:tau}) reduces to the known relation $\tau = 1 + d/\tilde{d}$ at $p_c$.
Figure~\ref{fig:sw-nsp} shows the scaling results for $n_{\geq}(s)$ at $p = 0.2$, $0.5$, and $0.8$.  
The data collapse onto a single curve for each value of $p$, confirming the validity of the scaling form.  
This implies that, in the large size limit, the cluster size distribution $n(s)$ follows a power law of the form $n(s) \propto s^{-\tau(p)}$.

We find that similar results hold even when the parameter $m$ (and hence the degree exponent $\gamma$) is varied (data not shown).  
As Eq.~(\ref{eq:psi-smallworld}) indicates, a nontrivial $p$-dependent fractal exponent exists for any value of $\gamma$, showing that the existence of a critical phase does not rely on degree heterogeneity.  

\begin{figure}[t]
\centering
(a) \includegraphics[height=5.5cm]{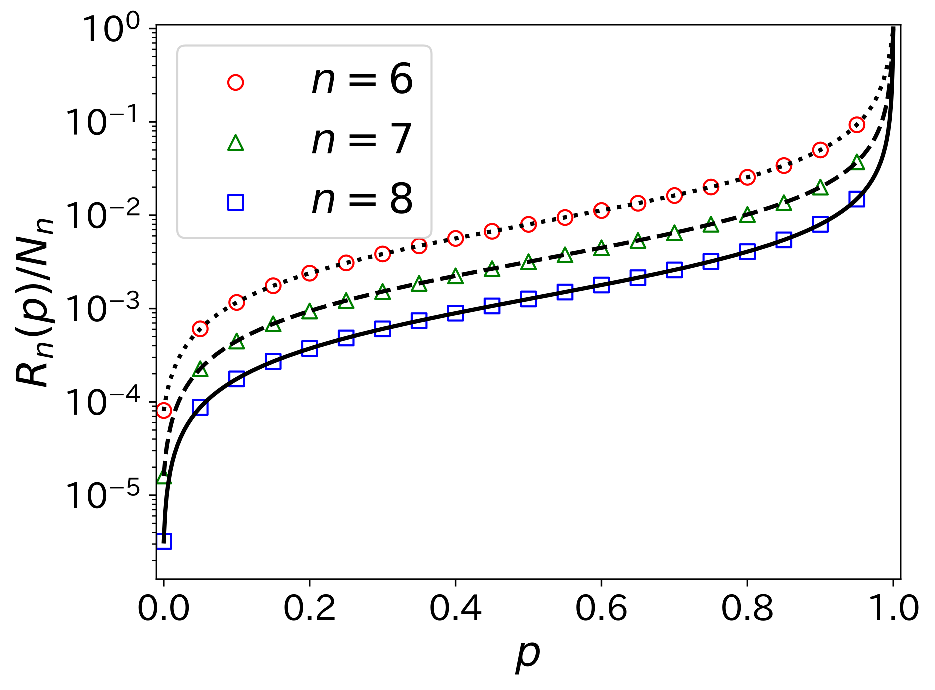}
(b) \includegraphics[height=5.5cm]{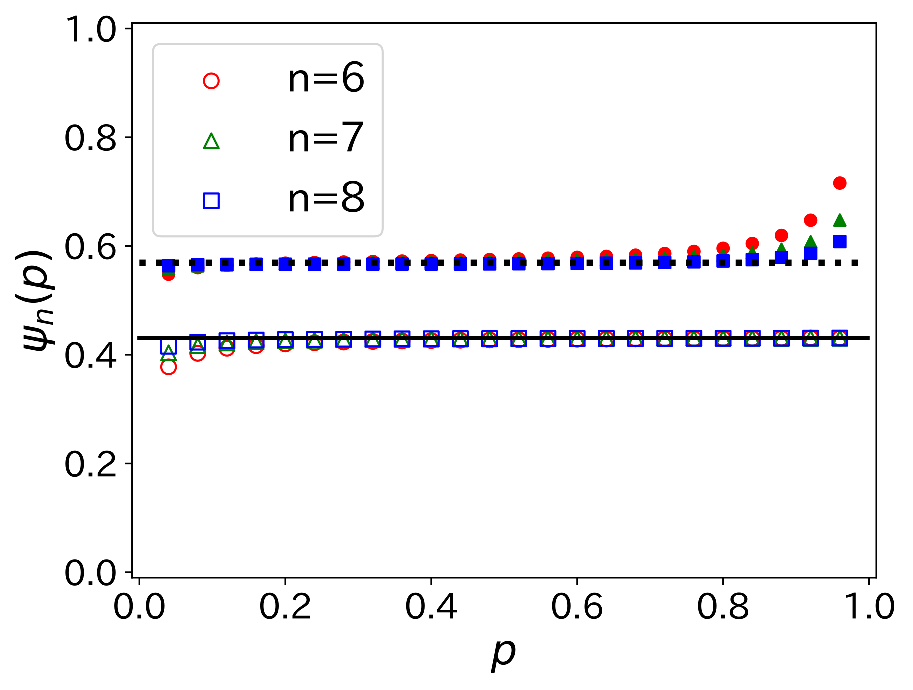}
\caption{
(a) $\bar{s}_{\mathrm{root}}(p) = R_n(p)/N_n$ and (b) $\psi_n(p)$ as a function of $p$ in the fractal SF tree.
Red open circles, green open triangles, and blue open squares represent Monte Carlo results for $G_n$ with $n=6$, $7$, and $8$, respectively.
In panel (a), the black dotted, dashed, and solid lines show the theoretical predictions for $R_n(p)/N_n$ of $G_6$, $G_7$, and $G_8$, respectively, obtained from Eq.~(\ref{eq:Nn}) and Eq.~(\ref{eq:root-fractal}).
In panel (b), the black solid line shows the theoretical curve of the fractal exponent $\psi (p)$, obtained from Eq.~(\ref{eq:psi-fractal}).
Panel (b) also includes results for the SHM network with $p_G=0.5$: red filled circles, green filled triangles, and blue filled squares represent Monte Carlo results of $\psi_n(p)$ for $n=6$, $7$, and $8$, respectively, and the black dotted line indicates the maximum degree exponent $\alpha$.
}\label{fig:mANDpsi-fractal}
\end{figure}

\begin{figure}[t]
\centering
(a) \includegraphics[height=5.5cm]{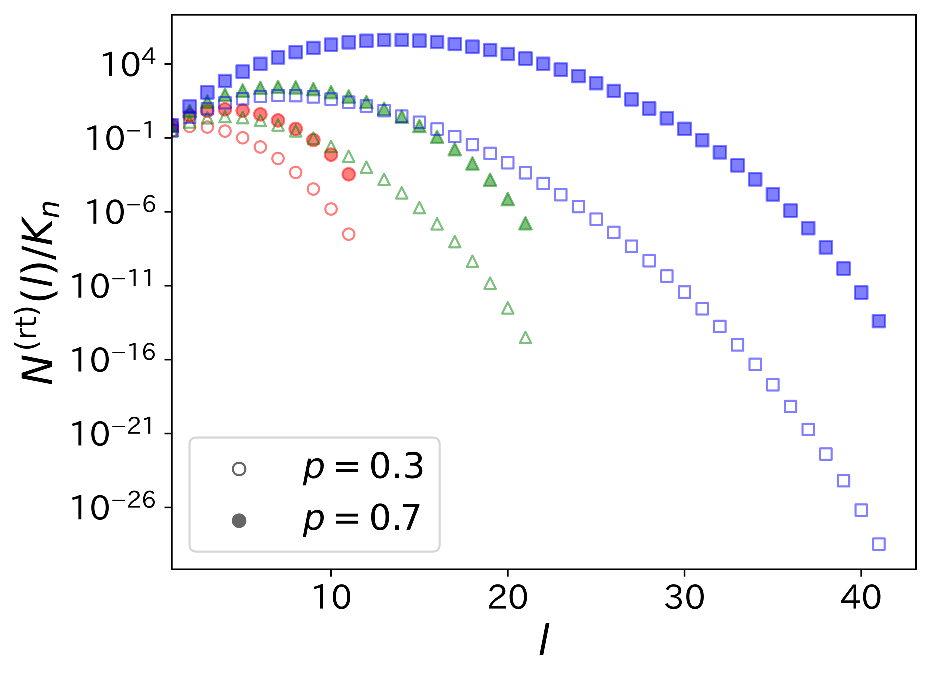}
(b) \includegraphics[height=5.5cm]{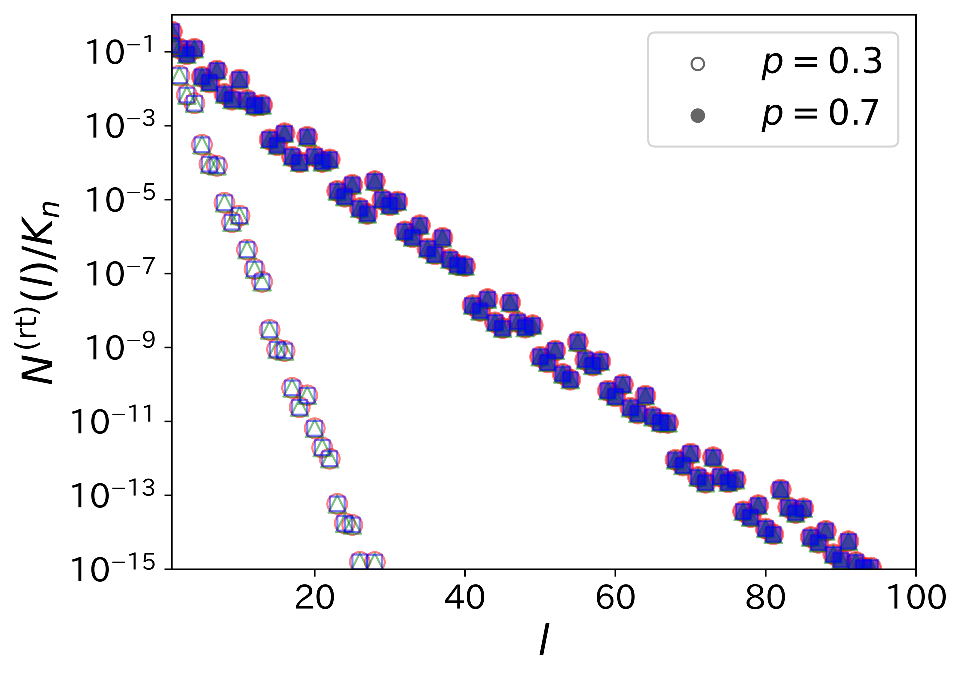}
\caption{
Distribution of $N_n^{\mathrm{(rt)}}(l)/K_n$ for (a) the small-world SF tree and (b) the fractal SF tree at $p = 0.3$ (open symbols) and $p = 0.7$ (filled symbols). 
Here, $N_n^{\mathrm{(rt)}}(l)$ is the number of nodes in the root cluster at distance $l$ from the root, and $K_n$ is the degree of the root. 
Red circles, green triangles, and blue squares represent the results for $G_n$ with $n = 10$, $20$, and $40$ in panel (a), and with $n = 6$, $8$, and $10$ in panel (b), respectively.
}\label{fig:correlationFunction}
\end{figure}

Next, we consider the fractal case.  
Figure~\ref{fig:mANDpsi-fractal} presents the order parameter $\bar{s}_{\mathrm{root}}(p)$ and the fractal exponent $\psi_n(p)$ for the fractal SF tree with $n = 6$, $7$, and $8$ as functions of $p$.  
As in the small-world case, $\bar{s}_{\mathrm{root}}(p)$ decreases with increasing $n$ and tends toward zero over the entire range of $p$ (Fig.~\ref{fig:mANDpsi-fractal}(a)), indicating the absence of a percolating phase ($p_{c2}=1$).
In contrast, the behavior of the fractal exponent differs from that in the small-world case.  
As shown in Fig.~\ref{fig:mANDpsi-fractal}(b), both the theoretical curve and the Monte Carlo results indicate that $\psi_n(p)$ converges to a constant value in the large size limit, specifically $\psi = \log 2 / \log 5$ for all $0 < p < 1$.

Although the fractal exponent $\psi(p)$ is nonzero, we conclude that bond percolation on the fractal SF tree remains in a nonpercolating phase for all $p < 1$, i.e., $p_{c1}=1$.  
This nonzero fractal exponent shown in Fig.~\ref{fig:mANDpsi-fractal}(b) can be attributed to the extremely large degree of the root.
To interpret this, we introduce another exponent $\alpha$, defined by the order of the maximum degree as $K_n \propto N_n^{\alpha}$ for $n \gg 1$.
From Eqs.~(\ref{eq:Nn}) and~(\ref{eq:Kn}), the exponent $\alpha$ for the SHM network with a given rewiring probability $p_G$ is given by
\begin{equation}
\alpha = \frac{ \log (m + 1 - p_G) }{ \log (2m + 1) }.  
\label{eq:alpha}
\end{equation}
As long as $p>0$, approximately $K_n p = O(N_n^\alpha)$ nodes become part of the root cluster solely due to direct connections to the root, which is connected to $K_n$ neighbors.
In fact, the fractal exponent derived earlier in Eq.~(\ref{eq:psi-fractal}) for the fractal SF tree is exactly equal to $\alpha$ with $p_G = 1$.

Figure~\ref{fig:correlationFunction} shows the number of nodes in the root cluster at distance $l$ from the root, $N_n^{\mathrm{(rt)}}(l)$, as a function of $l$, for several values of $p$, in both small-world and fractal SF  trees.  
In the small-world SF tree, $N_n^{\mathrm{(rt)}}(l)$ increases exponentially with $l$ over a range that expands with increasing $n$, whereas in the fractal SF tree, it decays rapidly with $l$.  
This indicates that although the root cluster in the fractal SF tree has a nonzero fractal exponent---due to the large number of nodes directly connected to the root---it does not extend far from the root node.
This observation supports the conclusion that the fractal SF tree remains in a nonpercolating phase for all $0 < p < 1$, despite exhibiting a nonzero fractal exponent.

Figure~\ref{fig:ns}(b) shows the cumulative cluster size distribution $n_{\geq}(s)$ for the fractal SF tree. 
Although $n_{\geq}(s)$ appears to follow a power-law form, this behavior merely reflects the intrinsic power-law nature of the degree distribution. 
The slope of $n(s)$ approximately matches the degree exponent $\gamma$ and remains unchanged across different values of $p$.

The above results remain unchanged even when $m$ is varied---that is, even when the fractal dimension and degree exponent of the fractal SF tree change. 
In $d$-dimensional Euclidean lattice systems, a percolating phase emerges when $d \ge 2$~\cite{stauffer2018introduction}.  
However, this criterion does not apply to trees.

As pointed out in Ref.~\cite{hasegawa2014critical}, on nonamenable graphs, the correlation length does not diverge at $p_{c1}$, but rather at $p_{c2}$.  
At $p_{c1}$, it is the correlation volume---the sum of correlation functions between a node and all other nodes---that diverges.  
In trees, which contain no cycles, the correlation function between two nodes at distance $l$---that is, the probability that the two nodes belong to the same cluster---decays exponentially as $p^l$.  
Therefore, the correlation length never diverges for any $p < 1$, and hence $p_{c2} = 1$.
On the other hand, if the number of reachable nodes from a given node increases exponentially with distance $l$ (as in the small-world case), this exponential growth can dominate the exponential decay of the correlation function.  
In such cases, the correlation volume can diverge, allowing for $p_{c1} < 1$.
However, in trees with finite fractal dimensions, the number of reachable nodes grows only polynomially (or subexponentially) with $l$, and the correlation volume does not diverge.  
Hence, $p_{c1} = 1$ in such cases.  
This conclusion holds as long as the tree retains a finite dimension.

Finally, we consider the intermediate case with $0 < p_G < 1$, in which edges are rewired with probability $p_G$.  
In this case, the percolation behavior is qualitatively similar to that of the fractal SF tree.  
Figure~\ref{fig:mANDpsi-fractal}(b) shows the fractal exponent of the SHM network with $p_G = 0.5$, obtained from Monte Carlo simulations.  
From the figure, we observe that in the large size limit, the fractal exponent remains constant and coincides with the exponent $\alpha$~(\ref{eq:alpha}) associated with the maximum degree.  
The cumulative cluster size distribution $n_{\geq}(s)$ also exhibits a power-law form, but its slope does not vary with $p$ (Fig.~\ref{fig:ns}(c)).  
Although both the degree heterogeneity and the fractal dimension vary with $p_G$ in the range $0 < p_G < 1$, the system remains in a nonpercolating phase throughout the entire range of $p$.

\section{Summary}

In this study, we investigated bond percolation on the Song--Havlin--Makse (SHM) network.  
This model allows the construction of both fractal scale-free (SF) trees with finite dimensionality and small-world SF trees with infinite dimensionality.  
Using a generating function approach, we formulated the average size of the root cluster and derived the corresponding fractal exponent for both types of trees.  
We found that when the SHM network is infinite-dimensional, bond percolation remains in the critical phase throughout the entire range of $p$ ($p_{c1}=0$ and $p_{c2}=1$).  
In contrast, when the network is finite-dimensional, the system remains in the nonpercolating phase for all $p$ ($p_{c1}=p_{c2}=1$).  
Although the order parameter (the fraction of the root cluster) vanishes in both cases, the behavior of the fractal exponent (the order of the size of the root cluster) differs markedly depending on the tree structure. 
In the infinite-dimensional cases, the fractal exponent varies continuously with $p$, indicating that the cluster size distribution follows a power law with a $p$-dependent slope.
In the finite-dimensional cases, the fractal exponent remains constant for all $p$.
This reflects that the root cluster arises primarily from the root's large degree and extends only a short distance.
These findings hold for the SHM network regardless of the value of the degree exponent $\gamma$ or its fractal dimension $d_{\mathrm{f}}$.

This study has demonstrated that dimensionality (not degree heterogeneity) is a decisive structural factor governing the possibility of a critical phase in tree networks. 
As long as the effective dimension of the tree remains finite, the correlation volume never diverges, and thus no critical phase can emerge.  
It is natural to expect that the same scenario holds even if the construction rule of the SHM network is modified, as long as it remains a tree.
The present results thus clarify the structural origin of critical phases and contribute to a broader understanding of universality in percolation on complex networks.
Nevertheless, infinite dimensionality alone is not sufficient to ensure the emergence of a critical phase in general networks.
For networks that are not trees, the conditions under which the correlation volume and the correlation length diverge at different values of $p$ ($p_{c1}<p_{c2}$) remain to be fully understood. 
These issues warrant further investigation.
Although this study focused on percolation, critical phases have also been reported in spin systems on complex networks~\cite{bauer2005phase, hinczewski2006inverted, boettcher2011fixed, boettcher2011renormalization, nogawa2012criticality, nogawa2012generalized, singh2014explosive, boettcher2015classification}.  
A natural extension of the present work is to study the Ising model on the SHM network, with particular emphasis on how the relaxation dynamics---namely, the time decay of the magnetization and the relaxation of the spin-spin autocorrelation function---differ between the small-world and fractal cases.
We leave this issue for future work.

\section*{Acknowledgment}

We thank Koji Nemoto and Jun Yamamoto for their useful comments and suggestions.
This work was supported by JSPS KAKENHI Grant Numbers JP21H03425 and JP24K06879.




\end{document}